\documentclass{article}
\usepackage{spconf,amsmath,graphicx}

\usepackage{algorithmic}
\usepackage{bm}
\usepackage{subcaption}
\usepackage{comment}
\usepackage{cite}

\usepackage{amsmath}
\usepackage{amssymb}
\usepackage{amsfonts}

\usepackage{threeparttable}

\usepackage{mathtools}

\title{Contextualized Automatic Speech Recognition \\with Dynamic Vocabulary}
\name{Yui Sudo$^1$, Yosuke Fukumoto$^1$, Muhammad Shakeel$^1$, Yifan Peng$^2$, Shinji Watanabe$^2$}
\address{
  $^1$Honda Research Institute Japan Co., Ltd., Japan\\
  $^2$Carnegie Mellon University, USA}
\begin{document}
%\ninept
%
\maketitle
\begin{abstract}
Deep biasing (DB) enhances the performance of end-to-end automatic speech recognition (E2E-ASR) models for rare words or contextual phrases using a bias list. However, most existing methods treat bias phrases as sequences of subwords in a predefined static vocabulary. 
This naive sequence decomposition produces unnatural token patterns, significantly lowering their occurrence probability.
More advanced techniques address this problem by expanding the vocabulary with additional modules, including the external language model shallow fusion or rescoring. However, they result in increasing the workload due to the additional modules.
This paper proposes a dynamic vocabulary where bias tokens can be added during inference. 
Each entry in a bias list is represented as a single token, unlike a sequence of existing subword tokens. This approach eliminates the need to learn subword dependencies within the bias phrases.
This method is easily applied to various architectures because it only expands the embedding and output layers in common E2E-ASR architectures. Experimental results demonstrate that the proposed method improves the bias phrase WER on English and Japanese datasets by 3.1 -- 4.9 points compared with the conventional DB method.
\end{abstract}

\section{Introduction}
\label{sec:intro}

End-to-end automatic speech recognition (E2E-ASR) \cite{prabhavalkar2023end,li2022recent} combines an acoustic model and a language model (LM) into a single neural network to improve ASR performance.
Various E2E-ASR methods have been proposed, including connectionist temporal classification (CTC) \cite{ctc1,ctc2}, recurrent neural network transducer (RNN-T) \cite{rnnt1,gulati2020conformer}, attention mechanisms \cite{chorowski2015attention,attention1,radford2023robust}, and their hybrids \cite{watanabe2017hybrid,Hu2020DeliberationMB,sudo20224d}. 
However, the effectiveness of E2E-ASR models strongly depends on the context of the training data. This can lead to performance inconsistencies in unseen user contexts, including named entities.
Retraining ASR models for every possible context is infeasible, highlighting the need for a method that allows users to efficiently contextualize models without additional training.

To address the challenge of contextualization in ASR, a common strategy is shallow fusion with an external LM \cite{Huang2020ClassLA,47384}. %,kojima2022}, 
This strategy frequently involves using a weighted finite state transducer (WFST) to create an in-class LM that improves the recognition of target named entities. 
In addition, %neural LM fusion methods 
\cite{Kannan2018AnAO,sriram2018cold,hori17_interspeech} attempt to improve accuracy by integrating an external neural LM with an E2E-ASR model. This integration involves rescoring the output hypotheses of the E2E-ASR model; however, additional training is required to integrate an external LM, which increases the workload.

Deep biasing (DB) \cite{deepcontext2018,Jain2020ContextualRF,huang2023contextualized,han2022improving,huber2021instant,zhou2023copyne,sudo2024_bias,shakeel2024_bias} efficiently contextualizes E2E-ASR models without retraining by using an editable list of phrases, referred to as a bias list. 
Many DB techniques use a cross-attention layer integrated to the middle of the E2E-ASR architecture to ensure accurate recognition of bias phrases. 
Previous studies \cite{huang2023contextualized,han2022improving,huber2021instant,zhou2023copyne,sudo2024_bias,shakeel2024_bias} have enhanced the effectiveness of the cross-attention layer with an auxiliary bias phrase detection loss function optimized by multitask learning. However, these cross-attention layers are tightly integrated into individual architectures (e.g., CTC, RNN-T, and attention), making their application to other architectures complex. Further, multitask learning requires multiple experimental training phases to fine-tune the learning weights, which is a time-intensive process.

In addition, most existing DB methods treat bias phrases as sequences of subwords in a predefined static vocabulary. 
This naive sequence decomposition produces unnatural token patterns, significantly lowering their occurrence probability.
For example, the personal name ``\textit{Nelly}'' can be segmented into a subword sequence, e.g., ``\textit{N}'', ``\textit{el}'', and ``\textit{ly}.'' However, if such token patterns are rare in the training data, their probability is significantly reduced.
To address this problem, several studies \cite{Le2021ContextualizedSE,qiu2023improving} have incorporated external text data to train an external LM or a text encoder.
However, this technique can considerably increase the workload. 
Other strategies have been proposed to improve contextualization using extra information, such as phonemes %for the bias phrases 
\cite{bruguier2019phoebe,chen2019joint,futami2023phoneme}, named entity tags \cite{sudo2023retraining}, %knowledge graph \cite{Das2022},
and synthesized speech \cite{wang2022towards,wang2023improving}; however, these strategies also increase the workload.

This paper proposes a simple but effective DB method by introducing dynamic vocabulary expansion where bias tokens can be added during inference. 
Each entry in a bias list is represented as a single token, unlike a sequence of existing subword tokens.
This approach bypasses the complex process of learning subword dependencies within bias phrases, enabling effective biasing without relying on external text data, unlike the previous methods \cite{Le2021ContextualizedSE,qiu2023improving}.
In addition, the proposed method is trained with a conventional E2E-ASR loss by dynamically expanding the vocabulary, eliminating the need for auxiliary losses, unlike the previous studies \cite{huang2023contextualized,han2022improving,huber2021instant,zhou2023copyne,sudo2024_bias}. 
Furthermore, compared with the previous methods \cite{han2022improving,huber2021instant,zhou2023copyne,sudo2024_bias,Le2021ContextualizedSE,qiu2023improving}, the proposed method can be more easily applied to various architectures because it only expands the embedding and output layers of common E2E-ASR models between CTC, RNN-T, and attention.
The main contributions of this study are as follows:
\begin{itemize}
\leftskip -5.5mm 
    \item We propose a simple but effective DB method based on the dynamic vocabulary.
    \item We verify that the proposed method performs well on both the Librispeech-960 dataset (English) and our in-house Japanese dataset.
    \item We demonstrate the effectiveness of the proposed method on various architectures, including CTC/attention-based offline systems and RNN-T-based streaming systems.
\end{itemize}

\section{End-to-end ASR}
\label{sec:Preliminary}

This section describes the E2E-ASR system, such as attention-based encoder-decoder and RNN-T, which is expanded to the proposed method. 
The following subsections describe the components of the E2E-ASR: the audio encoder and decoder.

\subsection{Audio encoder}
\label{sec:encoder}

The audio encoder comprises two convolutional layers, a linear projection layer, and $M_{\text{a}}$ conformer blocks \cite{gulati2020conformer}.
The convolutional layers subsample an audio feature sequence \begin{math}\bm{X}\end{math}, and the conformer blocks then transform the subsampled feature sequence to a $T$-length hidden state vector sequence $\bm{H} = [\bm{h}_1, \cdots , \bm{h}_T] \in \mathbb{R}^{d \times T}$ where $d$ represents the dimension:
\begin{equation}
\label{eq:audio}
    \bm{H} = \mathrm{AudioEnc}(\bm{X}).
\end{equation}
Each Conformer block has two feedforward layers, a multiheaded self-attention layer, a convolution layer, and a layer-normalization layer with residual connections.

\subsection{Decoder}
\label{sec:attention}

Given $\bm{H}$ generated by the audio encoder in Eq.~\eqref{eq:audio} and the previously estimated token sequence $y_{0:i-1} = [y_0, \cdots, y_{i-1}]$,
the decoder with  an embedding and output layer estimates the next token $y_i$ recursively as follows:
\begin{equation}
\label{eq:decoder}
    P(y_i|y_{0:i-1},\bm{X}) = \mathrm{Decoder}(y_{0:i-1}, \bm{H}),
\end{equation}
where $y_i$ is the $i$-th subword-level token in the pre-defined static vocabulary $\mathcal{V}^{\text{n}}$ of size $K$ ($y_i \in \mathcal{V}^{\text{n}}$).
Note that $\mathcal{V}^{\text{n}}$ contains the blank token $\phi$ for RNN-T-based systems. 

Specifically, the decoder comprises an embedding layer, a main decoder block (e.g., transformer blocks for attention-based systems and prediction/joint network for RNN-T-based systems), and an output layer.
First, the embedding layer with positional encoding converts the input non-blank token sequence $y_{0:i-1}$ to an embedding vector sequence $\bm{E}_{0:i-1} = [\bm{e}_0, \cdots, \bm{e}_{i-1}] \in \mathbb{R}^{d \times i}$ as follows:
\begin{equation}
    \bm{E}_{0:i-1} = \mathrm{Embedding}(y_{0:i-1}).
\label{eq:embedding}
\end{equation}
Thereafter, $\bm{E}_{0:i-1}$ is input to the main decoder block with the hidden state vectors $\bm{H}$ in Eq.~\eqref{eq:audio} to generate a hidden state vector $\bm{u}_i \in \mathbb{R}^{d}$ as follows:
\begin{equation}
    \bm{u}_{i} = \mathrm{MainBlock}(\bm{H}, \bm{E}_{0:i-1}).
\label{eq:main_decoder}
\end{equation}
For example, attention-based systems employ the transformer blocks, while RNN-T-based systems 
comprise a prediction network and a joint network for the main decoder block, respectively.
Subsequently, the output layer calculates the token-wise score $\bm{\alpha}^{\text{n}} = [\alpha^{\text{n}}_1, \cdots , \alpha^{\text{n}}_{K}]^T$ and the corresponding probability as follows:
\begin{equation}
    \bm{\alpha}^{\text{n}} = \mathrm{Linear}(\bm{u}_{i}),
\label{eq:score}
\end{equation}
\begin{equation}
\label{eq:normal_softmax}
    P \left(y_i \mid y_{0:i-1}, \bm{X}\right) = \mathrm{Softmax}(\bm{\alpha}^{\text{n}}).
\end{equation}
Here, the vocabulary size $K$ is pre-defined by a static token list.
By repeating these processes recursively, the posterior probability of the token sequence is formulated as follows:
\begin{equation}
\label{eq:posterior}
   P(Y \mid \bm{X}) =  \begin{dcases}
                    \prod_{i=1}^{S} P\left(y_{i} \mid y_{0:i-1}, \bm{X}\right) & (\text{attention}),\\
                    \sum_{Z \in \mathcal{B^{\text{-1}}}(Y)} P(Z \mid \bm{X}) & (\text{RNN-T}),
                \end{dcases}
\end{equation}
\vspace*{-2mm}
\begin{equation}
    P(Z \mid \bm{X}) = \prod_{i=1}^{T+S} P\left(y_i \mid y_{0:i-1}, \bm{X} \right).
\end{equation}
Here, the attention decoder directly outputs the $S$-length non-blank token sequence $Y = (y_i \mid i = 1, ..., S)$, 
while the RNN-T decoder outputs the ($T + S$)-length alignment sequence $Z = (y_i \mid i = 1, ..., T + S)$.
$\mathcal{B^{\text{-1}}}(Y)$ in the RNN-T-based systems is a set of all possible alignment sequences of $Y$. 
The model parameters are optimized by minimizing the negative log-likelihood described as follows: 
\begin{equation}
    L = - \log P(Y \mid \bm{X}).
\label{eq:loss_att}
\end{equation}
The embedding and output layers in Eqs.~\eqref{eq:embedding} and \eqref{eq:normal_softmax} are expanded to the proposed DB method in Section~\ref{sec:biasdecoder}.

\begin{figure*}[t!]
     \centering
     \hfill
     \begin{subfigure}[b]{0.45\linewidth}
         \centering
         \includegraphics[scale=0.38]{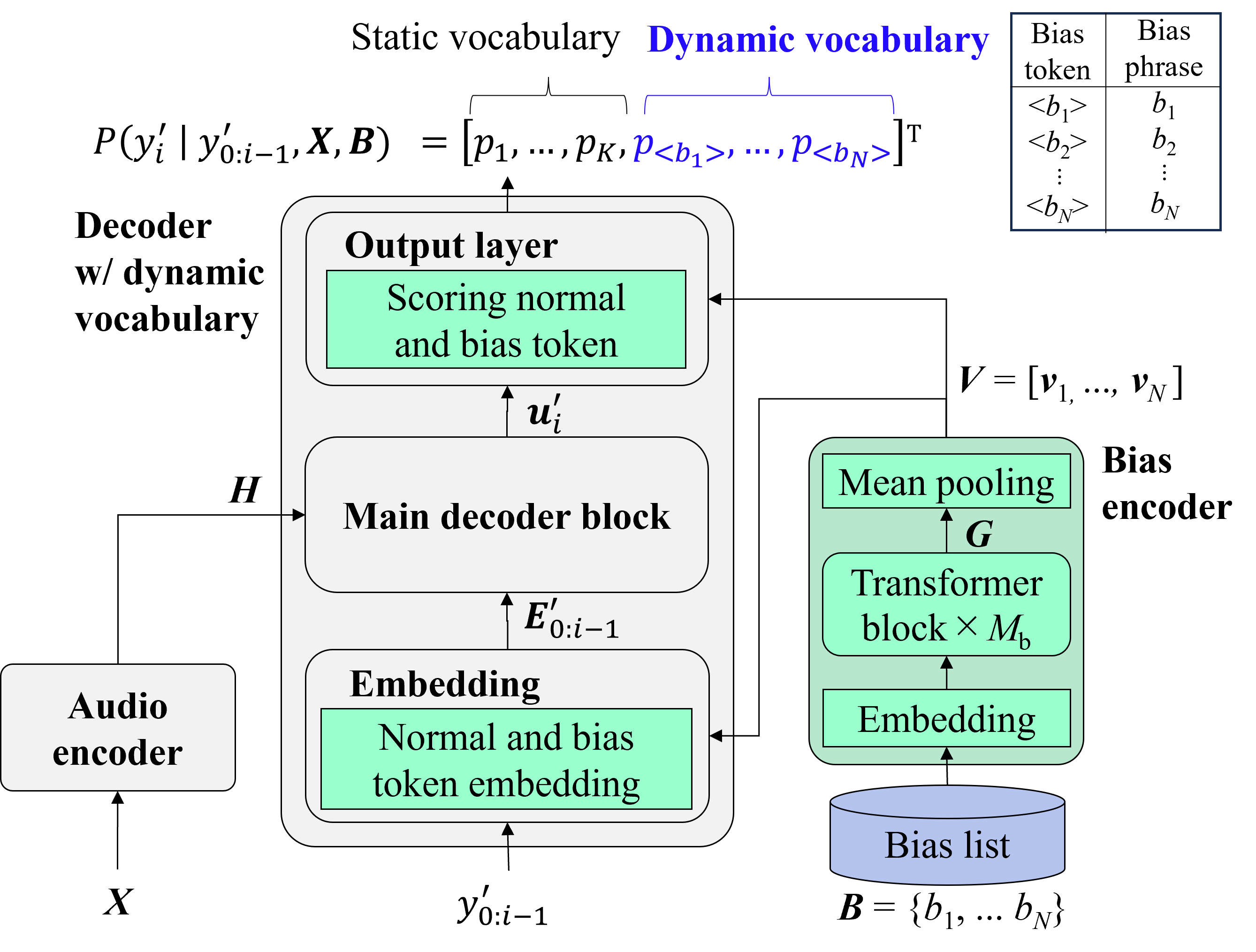}
         \vskip -0.0mm
         \caption{Overall architecture}
         \label{fig:overview}
     \end{subfigure}
     \hfill
     \begin{subfigure}[b]{0.27\linewidth}
         \centering
         \hskip 2.0mm
         \includegraphics[scale=0.4]{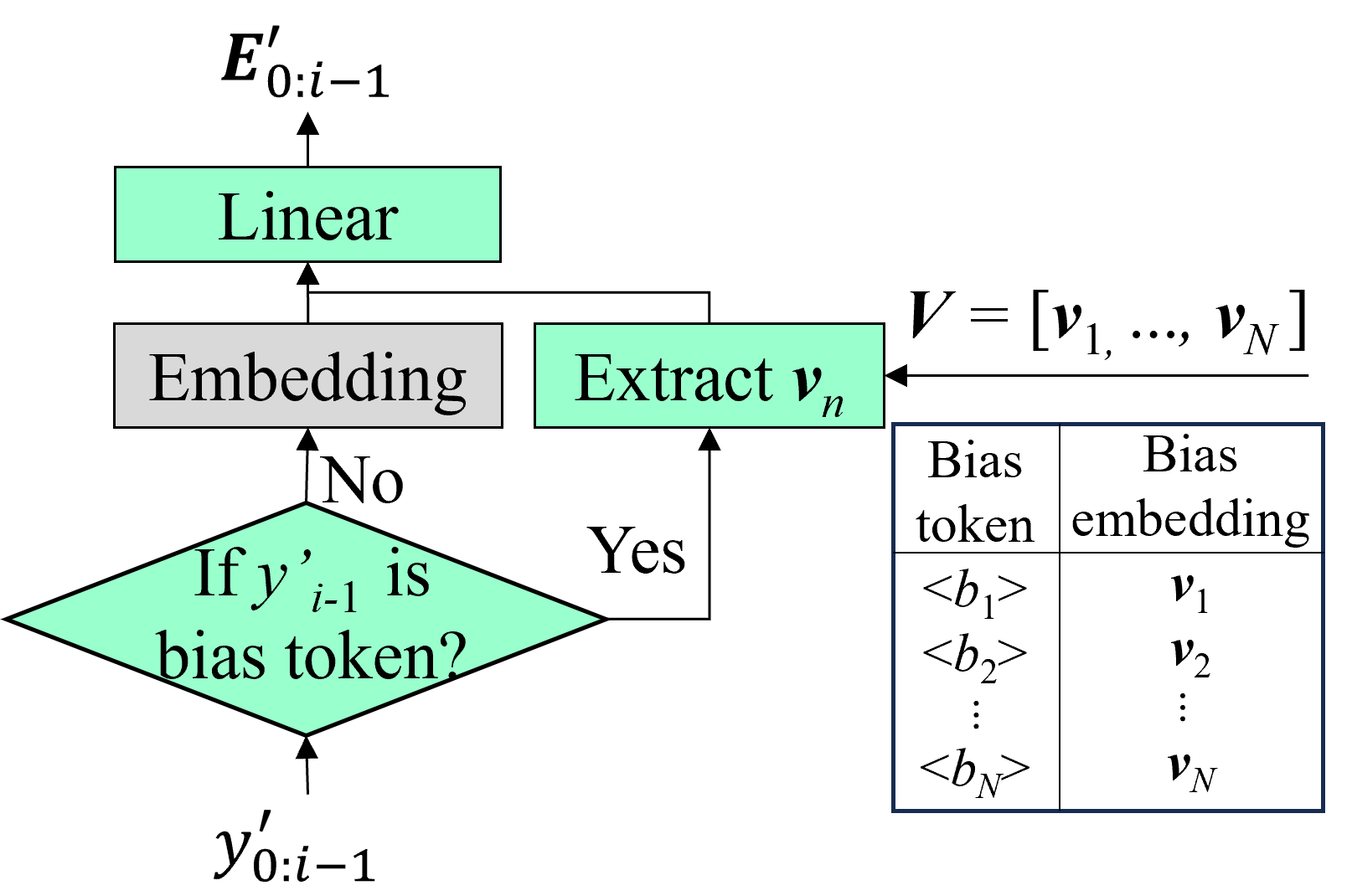}
         \vskip -0.0mm
         \caption{Expanded embedding layer}
         \label{fig:embeding}
     \end{subfigure}
     \hfill
     \begin{subfigure}[b]{0.27\linewidth}
         \centering
         \hskip 2.0mm
         \includegraphics[scale=0.4]{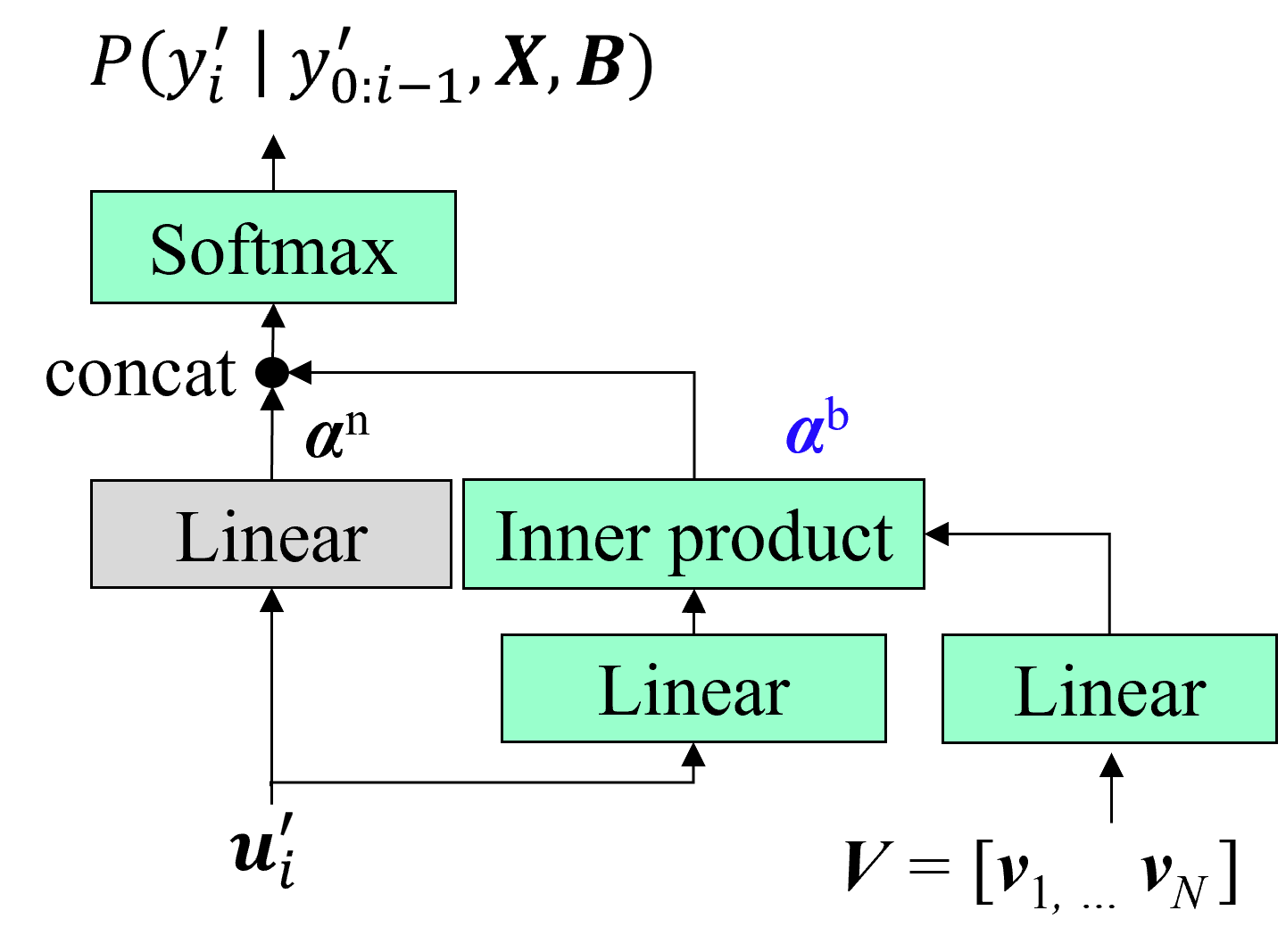}
         \vskip -0.0mm
         \caption{Expanded output layer}
         \label{fig:output}
     \end{subfigure}
     \hfill
    \vskip -6.0mm
    \caption{(a) Overall architecture of the proposed method, including the audio encoder, bias encoder, and decoder, with the expanded embedding and output layers. (b) Expanded embedding layer: If the input token is a dynamic bias token, the corresponding embedding $\bm{v}_n$ is extracted. (c) Expanded output layer: The bias score $\bm{\alpha}^{\text{b}}$ is calculated using the inner product.}
    \label{fig:component}
    \vskip -1.0mm
\end{figure*}

\section{Proposed method}
\label{sec:proposed}

Figure \ref{fig:component} shows the overall architecture of the proposed method, which comprises the existing audio encoder, as described in Section~\ref{sec:encoder}, newly introduced bias encoder, and decoder, which is nearly identical to Section~\ref{sec:biasdecoder} but with expanded embedding and output layers. The bias encoder and expanded decoder are described in the following subsections.

\subsection{Bias encoder}
\label{sec:biasencoder}

Similar to \cite{sudo2024_bias}, the bias encoder comprises an embedding layer with a positional encoding layer, 
$M_{\text{b}}$ transformer blocks, a mean pooling layer, and a bias list $B = \{b_{1}, \cdots, b_{N}$\}, where $b_{n} \in \mathcal{V}^{\text{n}}$ is the $I_n$-lengnth subword token sequence of the $n$-th bias phrase (e.g., [``\textit{N}'', ``\textit{el}'', ``\textit{ly}'']). 
After converting the bias list $B$ into a matrix $\bm{B} \in \mathbb{R}^{I_{\text{max}} \times N}$ through zero padding based on the maximum token length $I_{\text{max}}$ in $B$, the embedding layer and the $M_{\text{b}}$ transformer blocks extract the high level representation $\bm{G} \in \mathbb{R}^{d \times I_{\text{max}} \times N}$ as follows:
\begin{equation}
    \bm{G} = \mathrm{TransformerEnc}(\mathrm{Embedding}(\bm{B})).
\end{equation}
Then, a mean pooling layer extracts phrase-level embedding vectors $\bm{V} = [\bm{v}_{1}, \cdots, \bm{v}_{N}] \in \mathbb{R}^{d \times N}$ as follows:
\begin{equation}
    \bm{V} = \mathrm{MeanPool}(\bm{G}).
\label{eq:hp}
\end{equation}

\subsection{Expanded decoder with dynamic vocabulary}
\label{sec:biasdecoder}

To avoid the complexity associated with learning dependencies within the bias phrases,
we introduce a dynamic vocabulary $\mathcal{V}^{\text{b}} = \{<\hspace{-3pt}b_1\hspace{-3pt}>, \cdots, <\hspace{-3pt}b_N\hspace{-3pt}>\}$ where the phrase-level bias tokens represent the bias phrases with the single entities for the $N$ bias phrases in the bias list $\bm{B}$. 
Unlike Eq. \eqref{eq:decoder}, the expanded decoder estimates the next token $y_i^{\prime}$
from expanded vocabulary $\mathcal{V}^{\text{n}} \cup \mathcal{V}^{\text{b}}$, i.e., $y_i^{\prime} \in \mathcal{V}^{\text{n}} \cup \mathcal{V}^{\text{b}}$ given $\bm{H}$, $\bm{V}$ in Eqs. \eqref{eq:audio} and \eqref{eq:hp}, and $y^{\prime}_{0:i-1}$ as follows:
\begin{equation}
\label{eq:biasdecodeer}
    P(y^{\prime}_i | y^{\prime}_{0:i-1},\bm{X}, \bm{B}) = \mathrm{ExDecoder}(y^{\prime}_{0:i-1}, \bm{H}, \bm{V}),
\end{equation}
where $y^{\prime}_{0:i-1} = [y^{\prime}_0, \cdots , y^{\prime}_{i-1}]$ represents the expanded token sequence.
For example, if a bias phrase ``\textit{Nelly}'' exists in the bias list, the expanded decoder outputs the corresponding bias token [$<$\textit{Nelly}$>$] rather than the decomposed normal token sequence [``\textit{N}'', ``\textit{el}'', ``\textit{ly}''].

Similar to the conventional decoder described in Section \ref{sec:attention}, 
the decoder comprises an expanded embedding layer, a main decoder block, and an expanded output layer.
First, the input token sequence $y^{\prime}_{0:i-1}$ is converted into the embedding vector sequence $\bm{E}^{\prime}_{0:i-1} = [\bm{e}^{\prime}_0, \cdots , \bm{e}^{\prime}_{i-1}] \in \mathbb{R}^{d \times i}$.
Unlike Eq.~\eqref{eq:embedding}, if the input token $y^{\prime}_{i-1}$ is a bias token, the corresponding bias embedding $\bm{v}_{n}$ is extracted from $\bm{V}$ (Figure \ref{fig:embeding}); otherwise, the normal embedding layer is used with a linear layer as follows:
\begin{equation}
\label{eq:extended_embed}
   \bm{e}^{\prime}_{i-1} = \begin{cases}
                    \mathrm{Linear}(\mathrm{Embedding}(y^{\prime}_{i-1}))       & (y^{\prime}_{i-1} \in \mathcal{V}^{\text{n}})\\
                    \mathrm{Linear}(\mathrm{Extract}(\bm{V}, y^{\prime}_{i-1})) & (y^{\prime}_{i-1} \in \mathcal{V}^{\text{b}}).
                \end{cases}
\end{equation}
Subsequently, the main decoder block converts $\bm{E}^{\prime}_{0:s-1}$ into the hidden state vector $\bm{u}^{\prime}_{s}$ as in Eq.~\eqref{eq:main_decoder}. 
In addition to the normal token score $\bm{\alpha}^{\text{n}} = [\alpha^{\text{n}}_1, \cdots , \alpha^{\text{n}}_{K}]^T$ in Eq. \eqref{eq:score}, the bias token score $\bm{\alpha}^{\text{b}} = [\alpha^{\text{b}}_1, \cdots, \alpha^{\text{b}}_{N}]^T$ is calculated using an inner product with two linear layers (Figure \ref{fig:output}) as follows:
\begin{align}
    \label{eq:inner}
    & \bm{\alpha}^{\text{b}} = \frac{\mathrm{Linear}(\bm{u}^{\prime}_{i}) \mathrm{Linear}(\bm{V}^{T})}{\sqrt{d}}.
\end{align}
By concatenating the normal token score $\bm{\alpha}^{\text{n}}$ with the bias token score $\bm{\alpha}^{\text{b}}$, which results in $\bm{\alpha} = [\alpha^{\text{n}}_1, \cdots, \alpha^{\text{n}}_{K}, \alpha^{\text{b}}_1, \cdots, \alpha^{\text{b}}_{N}]^T$, Eq.~\eqref{eq:normal_softmax} can be expanded as follows:
\begin{equation}
\hspace*{-1mm}
    P\left(y^{\prime}_{i} \mid y^{\prime}_{0:i-1}, \bm{X}, \bm{B} \right) = \mathrm{Softmax}(\mathrm{Concat}(\bm{\alpha}^{\text{n}}, \bm{\alpha}^{\text{b}})).
\label{eq:softmax}
\end{equation}
%The computational cost of Eqs. (14) and (15) are equivalent to a single cross-attention layer in other DB methods. Since other DB methods introduce multiple cross-attention layers [24-27], the proposed method results in a smaller increase in computational cost.
Similar to Eqs.~\eqref{eq:posterior} -- \eqref{eq:loss_att}, the posterior probability and the loss function are formulated as follows:
\begin{equation}
\label{biaslikelihood}
   P(Y^{\prime} \mid \bm{X}, \bm{B}) =  \begin{dcases}
                    \prod_{i=1}^{S^{\prime}} P\left(y^{\prime}_{i} \mid y^{\prime}_{0:i-1}, \bm{X}, \bm{B}\right) & (\text{attention}),\\
                    \sum_{Z^{\prime} \in \mathcal{B^{\text{-1}}}(Y^{\prime})} P(Z^{\prime} \mid \bm{X}, \bm{B}) & (\text{RNN-T}),
                \end{dcases}
\end{equation}
\vspace*{-2mm}
\begin{equation}
    P(Z^{\prime} \mid \bm{X}, \bm{B}) = \prod_{i=1}^{T+S^{\prime}} P\left(y^{\prime}_i \mid y^{\prime}_{0:i-1}, \bm{X}, \bm{B} \right),
\end{equation}
\begin{equation}
    \label{eq:loss_bias}
    L^{\prime} = - \log P(Y^{\prime} \mid \bm{X}, \bm{B}),
\end{equation}
where $Y^{\prime}$ and $Z^{\prime}$ represent the $S^{\prime}$-length non-blank token sequence and ($T + S^{\prime}$)-length alignment sequence based on the proposed dynamic vocabulary, respectively.
Note that Eqs. \eqref{eq:inner} and \eqref{eq:softmax} do not hold learnable parameters depending on the bias list size $N$, and can replace the bias list dynamically during inference. 
Also, the proposed method is optimized only with Eq. \eqref{eq:loss_bias} without the auxiliary loss.

\begin{figure}[t!]
     \centering
     \hfill
     \begin{subfigure}[b]{0.35\linewidth}
         \centering
         \includegraphics[scale=0.33]{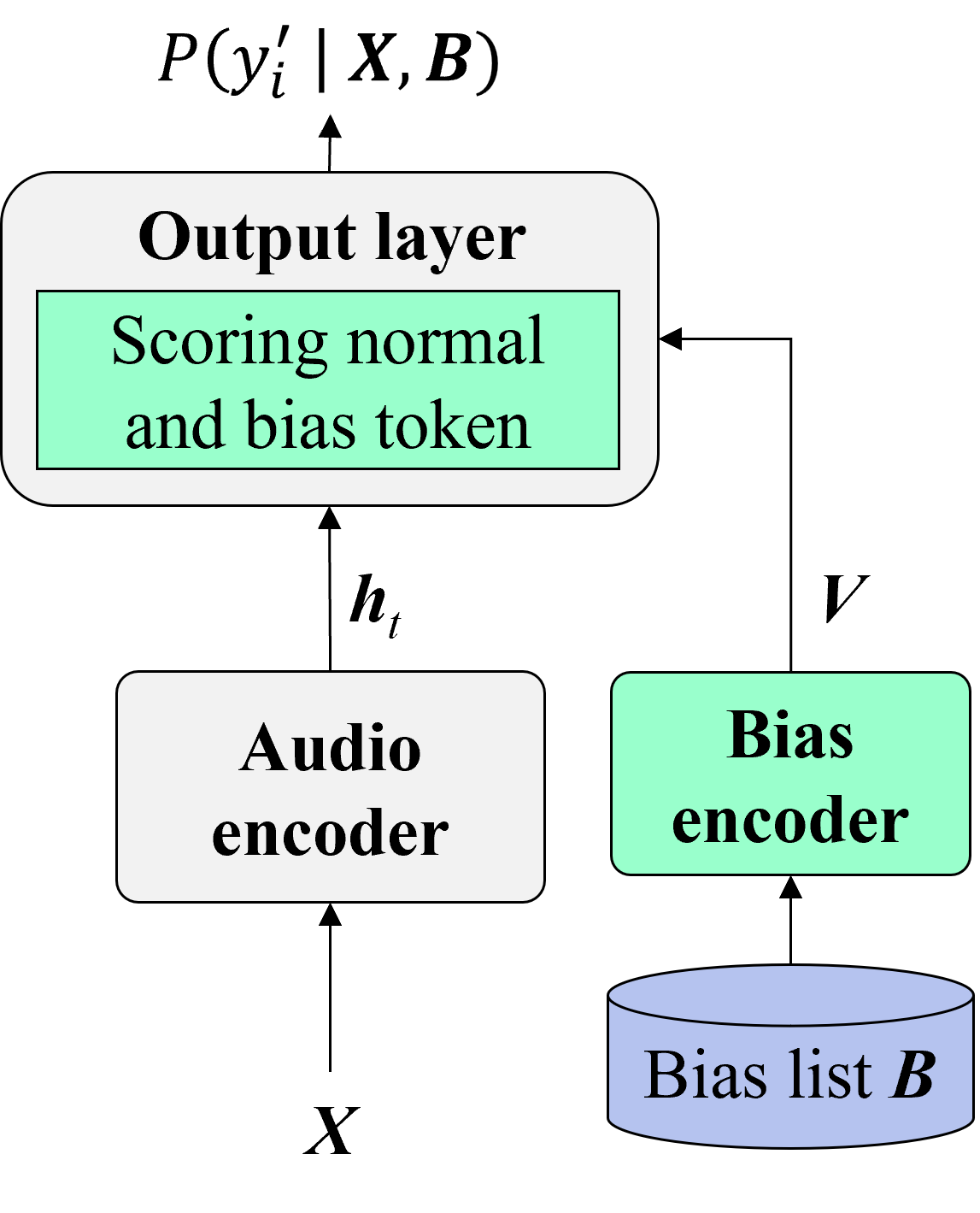}
         \vskip -0.0mm
         \caption{CTC-based system}
     \end{subfigure}
     \hfill
     \begin{subfigure}[b]{0.64\linewidth}
         \centering
         \hskip 2.5mm
         \includegraphics[scale=0.33]{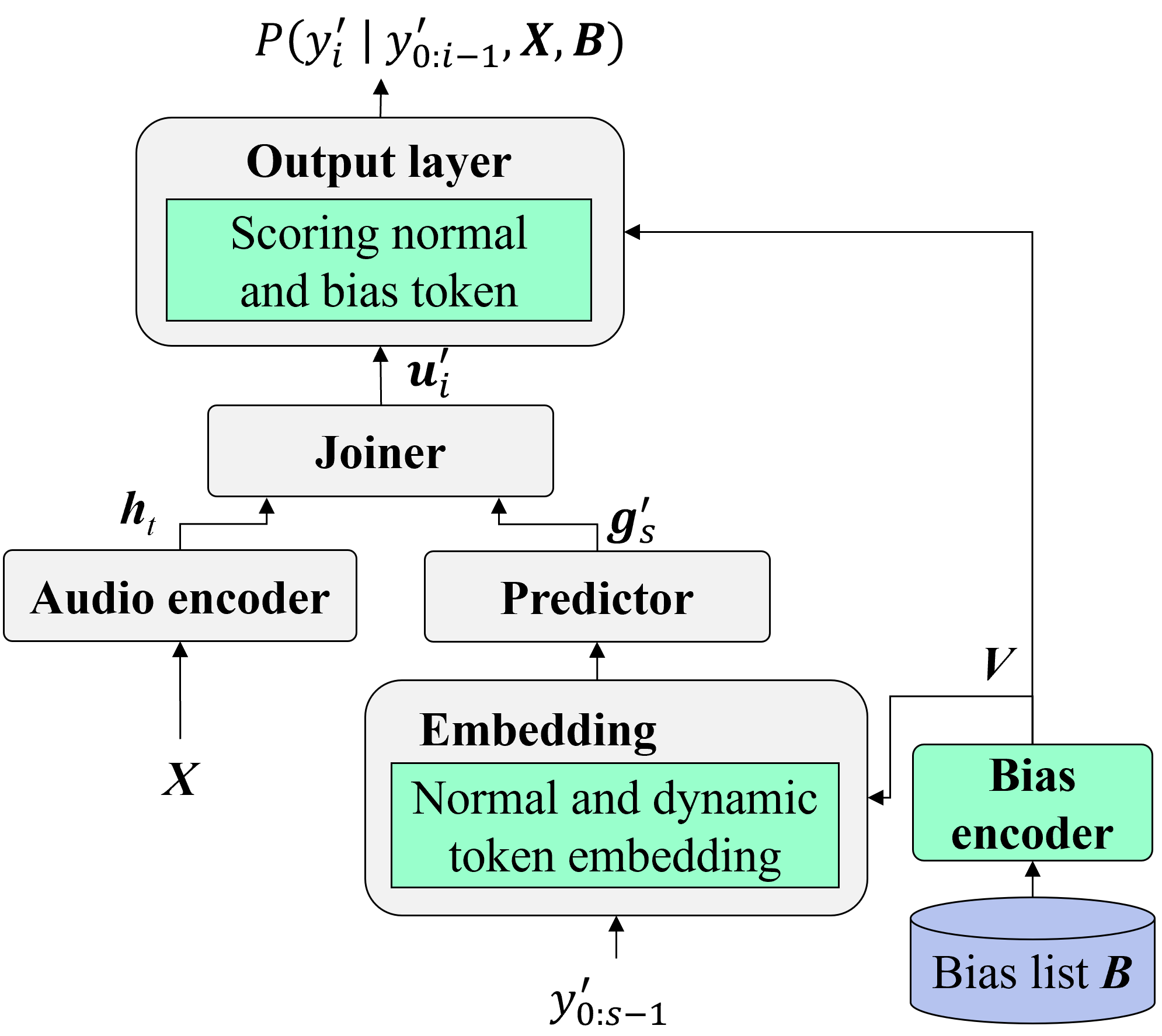}
         \vskip -0.0mm
         \caption{RNN-T-based system}
     \end{subfigure}
     \hfill
    \vskip -5.0mm
    \caption{Various architectures utilized in the proposed method.}
    \label{fig:transducer}
    \vskip -1.0mm
\end{figure}

The proposed method can be easily applied to various E2E-ASR architectures (e.g., CTC, RNN-T, and attention), including streaming systems and multilingual systems \cite{ctc2,rnnt1,radford2023robust,peng2024owsmctc} without major modifications, because it only expands the embedding and output layers in addition to the bias encoder (Figure~\ref{fig:transducer}). Note that since CTC does not have the embedding layer and the main decoder block, only the output layer is expanded as described in Eqs. \eqref{eq:inner} and \eqref{eq:softmax} using the hidden state vector $\bm{h}_t$ instead of $\bm{u}_i$ (Figure~\ref{fig:transducer}a).

%The proposed method allows the system to be flexibly determined depending on the application scenario. %For example, the proposed method can be applied to an RNN-T-based system for streaming systems, while hybrid CTC/attention is suitable for offline systems.

\subsection{Application to hybrid E2E-ASR systems}
\label{sec:dv-based-rnnt}

Given the simplicity of the proposed method, the proposed method can also be applied to hybrid systems, such as \cite{watanabe2017hybrid,wang2023accelerating,Hu2020DeliberationMB,sudo20224d,peng2023reproducing,peng2024owsm}, by expanding the output layer of each branch.
In this paper, the attention-based and RNN-T-based dynamic vocabulary models described in Sections~\ref{sec:biasdecoder} are trained with an auxiliary CTC loss, which is also based on the dynamic vocabulary, with the training weight $\lambda$ as follows:
\begin{equation}
\label{eq:lambda}
    L^{\prime}_{\text{joint}} = (1 - \lambda) L^{\prime} + \lambda L^{\prime}_{\text{ctc}},
\end{equation}
where $L^{\prime}_{\text{joint}}$ and $L^{\prime}_{\text{ctc}}$ represent loss functions for the joint model and auxiliary CTC decoder, respectively.

Moreover, the flexibility of the proposed method is preserved in joint decoding with multiple decoders \cite{watanabe2017hybrid,Hu2020DeliberationMB,sudo20224d,sudo20244d,sudo23c_interspeech,tsunoo23_interspeech}. %, because it only expands the vocabulary using the expanded embedding and output layers.
We adopt the joint decoding algorithms similar to \cite{watanabe2017hybrid,sudo20224d}. Specifically, the primary decoder (i.e., attention or RNN-T) generate the hypotheses and the scores of the hypotheses are augmented by the CTC decoder with the decoding weight $\gamma$ as follows:
\begin{align}\begin{split}
\label{eq:gamma}
     \beta_{\text{joint}} = (1 - \gamma) \beta_{\text{}} + \gamma \beta_{\text{ctc}},
\end{split}
\end{align}
\begin{equation}
   \begin{cases}
                    \beta_{\text{}} = \text{log} P(Y^{\prime} \mid \bm{X}, \bm{B})  & (\text{attention/RNN-T})\\
                    \beta_{\text{ctc}} = \text{log} P_{\text{ctc}}(Y^{\prime} \mid \bm{X}, \bm{B}) & (\text{CTC}),
                \end{cases}
\end{equation}
where $\beta_{\text{joint}}$, $\beta_{\text{}}$, and $\beta_{\text{ctc}}$ represent the scores of joint decoding, primary decoder, and CTC decoder, respectively.

\subsection{Training}
\label{sec:training}

During training, a bias list $\bm{B}$ is created randomly from the reference transcriptions for each batch, where $N_{\text{utt}}$ bias phrases are selected per utterance, each having a token length of $I$.
This process yields a total of $N$ bias phrases calculated as $N_{\text{utt}} \times$ batch size.
Once the bias list $\bm{B}$ is defined, the corresponding reference transcription $y_{\text{gt}}$ is modified to $y^{\prime}_{\text{gt}}$ based on the dynamic vocabulary.
For example, if the phrase ``\textit{N}'', ``\textit{el}'', ``\textit{ly}'' ($N_{\text{utt}}=1, I=3$) is extracted as a bias phrase from the reference transcription $y_{\text{gt}}$ =  [``\textit{Hi}'', ``\textit{N}'', ``\textit{el}'', ``\textit{ly}''], the reference transcription is modified to $y^{\prime}_{\text{gt}}$ =  [``\textit{Hi}'', $<$\textit{Nelly}$>$].

\subsection{Bias weight during inference}
\label{sec:decoding}

Considering practicality, we introduce a bias weight to Eq.~\eqref{eq:softmax} to avoid over/under-biasing during inference:
\begin{equation}
    \mathrm{WeightSoftmax}_{j}(\bm{\alpha}, \bm{w}) = \frac{w_j \mathrm{exp}(\alpha_j)}{\sum_{l=1}^{(K+N)} w_l \mathrm{exp}(\alpha_l)},
\label{eq:weight}
\end{equation}
where $\bm{w} = [w_1. \cdots, w_{(K+N)}]^T$ and $j$ represent a weight vector and its index for $\bm{\alpha} = [\alpha^{\text{n}}_1, \cdots , \alpha^{\text{n}}_{K}, \alpha^{\text{b}}_1, \cdots , \alpha^{\text{b}}_{N}]^T$, respectively.
The same bias weight $\mu$ is applied to the bias tokens as follows:
\begin{equation}
\label{eq:mu}
    w_j = \begin{cases}
                    1.0 & (j \leqq K)\\
                    \mu   & (j > K),
                \end{cases}
\end{equation}
if $\mu < 1.0$, the bias tokens are underweighted compared to the normal tokens; otherwise, the bias tokens are overweighted compared to the normal tokens.
%Note that $\mu$ may be greater or less than 1.0.

\begin{table*}[h]
\caption{WER results of the \textbf{offline CTC/attention-based} systems obtained on Librispeech-960 (U-WER/B-WER).}
\vspace*{-7mm}
\label{maintable}
\begin{center}
\resizebox {0.98\linewidth} {!} {
\begin{tabular}{@{}c|cc|cc|cc|cc}
\hline
      & \multicolumn{2}{c|}{$N$ = 0 (no-bias)} &\multicolumn{2}{c|}{$N$ = 100} & \multicolumn{2}{c|}{$N$ = 500} & \multicolumn{2}{c}{$N$ = 1000} \\% & \multicolumn{2}{c}{$N$ = 2000}  \\
Model & test-clean & test-other & test-clean & test-other & test-clean & test-other & test-clean & test-other \\%& test-clean & test-other  \\
\hline
Baseline  & \textbf{2.57} & \textbf{5.98} & 2.57 & 5.98 & 2.57 & 5.98 & 2.57 & 5.98 \\
(CTC/attention) & (\textbf{1.5}/\textbf{10.9}) & (\textbf{4.0}/\textbf{23.1}) & (\textbf{1.5}/10.9) & (\textbf{4.0}/23.1) & (\textbf{1.5}/10.9) & (\textbf{4.0}/23.1) & (\textbf{1.5}/10.9) & (\textbf{4.0}/23.1) \\
\hline
CPPNet \cite{huang2023contextualized} & 4.29 & 9.16  & 3.40 & 7.77 & 3.68 & 8.31 & 3.81 & 8.75 \\%& N/A & N/A \\
 & (2.6/18.3) & (5.9/37.5) & (2.6/10.4) & (6.0/23.0) & (2.8/10.9) & (6.5/24.3) & (2.9/11.4) & (6.9/25.3) \\%& N/A & N/A \\
 \hline
Attention-based DB & 5.05 & 8.81 & 2.75 & 5.60 & 3.21 & 6.28 & 3.47 & 7.34 \\
+ BPB beam search \cite{sudo2024_bias} & (3.9/14.1) & (6.6/27.9) & (2.3/6.0) & (4.9/12.0) & (2.7/7.0) & (5.5/13.5) & (3.0/7.7) & (6.4/15.8) \\
\hline
Proposed & 3.16 & 6.95 & \textbf{1.80} & \textbf{4.63} & \textbf{1.92} & \textbf{4.81} & \textbf{2.01} & \textbf{4.97} \\%& 2.17 & 5.19 \\ 
 & (1.9/13.8) & (4.6/27.5) & (1.7/\textbf{2.8}) & (4.3/\textbf{7.1}) & (1.8/\textbf{3.1}) & (4.5/\textbf{7.9}) & (1.9/\textbf{3.3}) & (4.6/\textbf{8.5}) \\%& (3.0/\textbf{2.6}) & (4.8/\textbf{8.9}) \\
\hline
\end{tabular}
}
\end{center}
\vspace*{-5mm}
\end{table*}

\section{Experiment}
\vspace*{-1mm}
To verify the effectiveness of the proposed method, we apply it to offline CTC/attention %\cite{watanabe2017hybrid} 
and streaming RNN-T models.

\vspace*{-2mm}
\subsection{Experimental setup}
\vspace*{-1mm}
\label{sec:experimental condition}

The input features are 80-dimensional Mel filterbanks with a window size of 512 samples and a hop length of 160 samples. Subsequently, SpecAugment %\cite{specaug} 
is applied.
The audio encoder comprises two convolutional layers with a stride of two and a 256-dimensional linear projection layer followed by 12 conformer layers with 1024 linear units and layer normalization. 
For the streaming RNN-T model, the audio encoder is processed blockwisely \cite{streaming-transformer-encoder} with block size and look ahead of 800 and 320 ms, respectively.
The bias encoder has six transformer blocks with 1024 linear units.
Regarding the expanded decoder, the offline CTC/attention model has six transformer blocks with 2048 linear units, and the streaming RNN-T model has a single long short-term memory layer with a hidden size of 256 and a linear layer of 320 joint sizes for prediction and joint networks.
The attention layers in the audio encoder, bias encoder, and expanded decoder are four multihead attentions with a dimension $d$ of 256. 

The offline CTC/attention and streaming RNN-T models have 40.58 M and 31.38 M parameters, respectively, including the bias encoders.
The training weight $\lambda$ in Eq.~\eqref{eq:lambda} is 0.3 for the CTC/attention and RNN-T models. The decoding weight $\gamma$ in Eq.~\eqref{eq:gamma} is 0.3 and 0.1 for the CTC/attention and RNN-T models, respectively. 
The bias weight $\mu$ in Eq.~\eqref{eq:mu} is set to 0.8 and 0.01 for the CTC/attention and RNN-T models, respectively (this is discussed further in Section \ref{sec:weight}).
During training, a bias list $\bm{B}$ is created randomly for each batch with $N_{\text{utt}}$ = [2 - 10] and $I$ = [2 - 10] (Section~\ref{sec:training}). 
The proposed models are trained for 150 epochs at a learning rate of 0.0025 and 0.002 for the CTC/attention-based and RNN-T-based systems, respectively.

The Librispeech-960 corpus \cite{panayotov2015librispeech} is employed to evaluate the proposed method using ESPnet toolkit \cite{espnet}. 
The proposed method is evaluated in terms of the word error rate (WER), bias phrase WER (B-WER), and unbiased phrase WER (U-WER) as in \cite{Le2021ContextualizedSE}.
The static vocabulary size $K$ is 5000, while the dynamic vocabulary size $N$ ranges from 0 to 2000.

\subsection{Results of the offline CTC/attention-based system}

Table \ref{maintable} shows the results of the offline CTC/attention-based systems obtained on the Librispeech-960 dataset for different bias list sizes $N$. 
With a bias list size of $N > 0$, the proposed method improves the B-WER considerably despite a slight increase in the U-WER, resulting in a substantial improvement in the overall WER. 
While the B-WER and U-WER tend to deteriorate with larger $N$, the proposed method remains superior to other DB techniques across all bias list sizes. 
In addition, the proposed method shows significant B-WER improvement for unseen words in the training data. Specifically, the baseline B-WER for unseen words in the test-other set is 73.5\%, whereas the proposed method improves the B-WER to 19.0\% when the bias list size is $N=1000$.
%Although the proposed method does not perform as well as the baseline when $N = 0$, this is not considered a particularly limiting drawback considering that users typically register important keywords in the bias list.

\subsection{Analysis of the proposed bias token}
\label{sec:analysis}

\begin{figure}[t]
        \begin{minipage}{0.49\textwidth}
        \centering
            \includegraphics[width=0.8\textwidth]{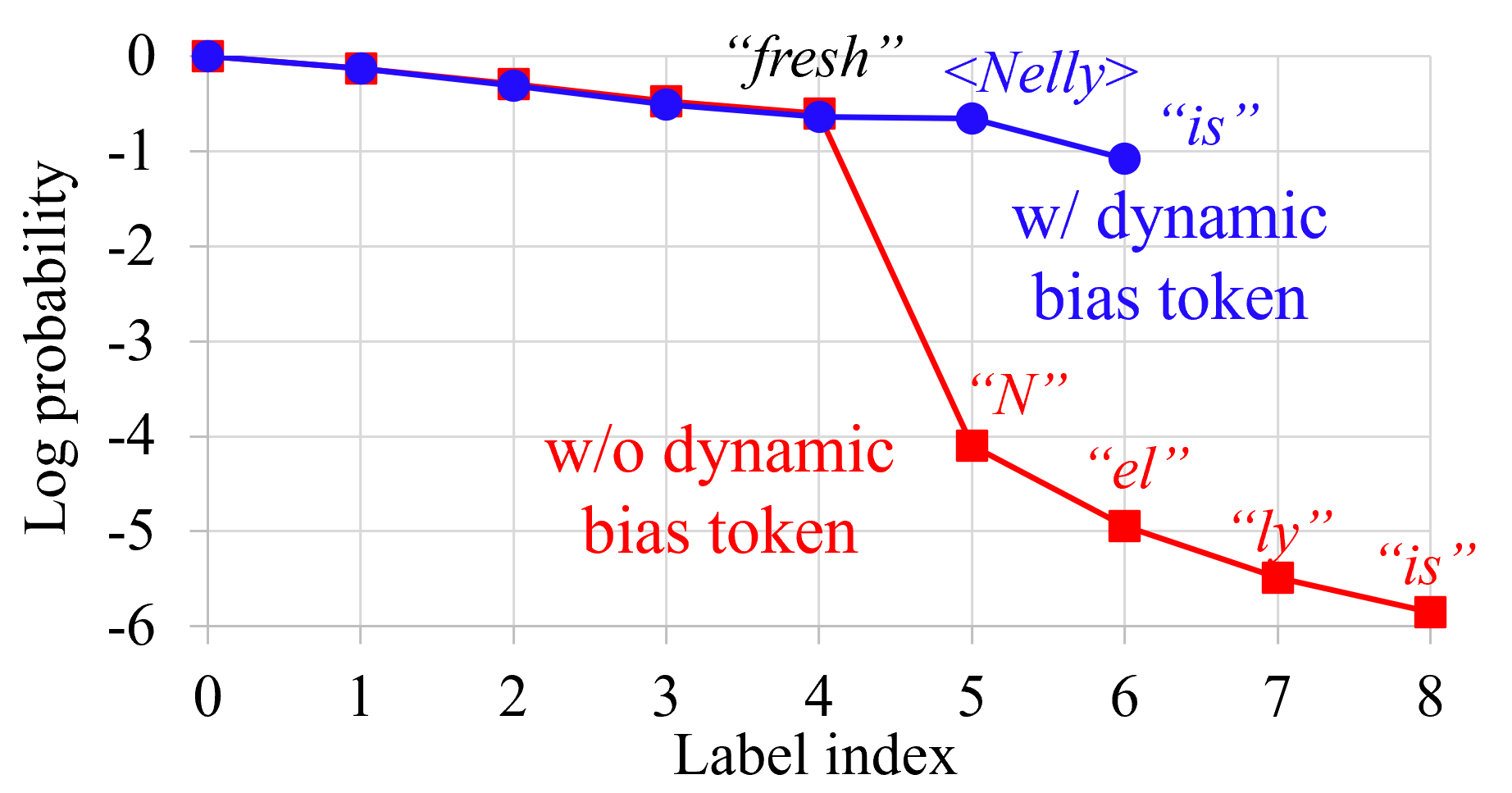} 
        \end{minipage}
    \vspace*{-3mm}
    \caption{Example of cumulative log probability during beam search.}
    \label{fig:prob}
\vspace*{-3mm}
\end{figure}

Figure \ref{fig:prob} shows an example of the cumulative log probability described in Eq.~\eqref{biaslikelihood}, where the blue and red lines indicate the results obtained \textit{with} and \textit{without} the bias tokens. 
Without using the bias tokens, the model struggles to capture the subword dependencies, resulting in significantly lower scores for each subword. 
Conversely, the proposed method assigns a high score to the bias token ($<$\textit{Nelly}$>$), improving the B-WER (Table \ref{maintable}). 
Interestingly, the log probabilities before and after the bias token (``\textit{fresh}'' and ``\textit{is}'') remain stable, even though the bias tokens are created dynamically during inference. 
This indicates that the proposed method preserves the context with non-bias tokens while eliminating the need to learn subword dependencies within the bias phrases.

\begin{figure}[t]
        \begin{minipage}{0.49\textwidth}
        \centering
            \includegraphics[width=0.75\textwidth]{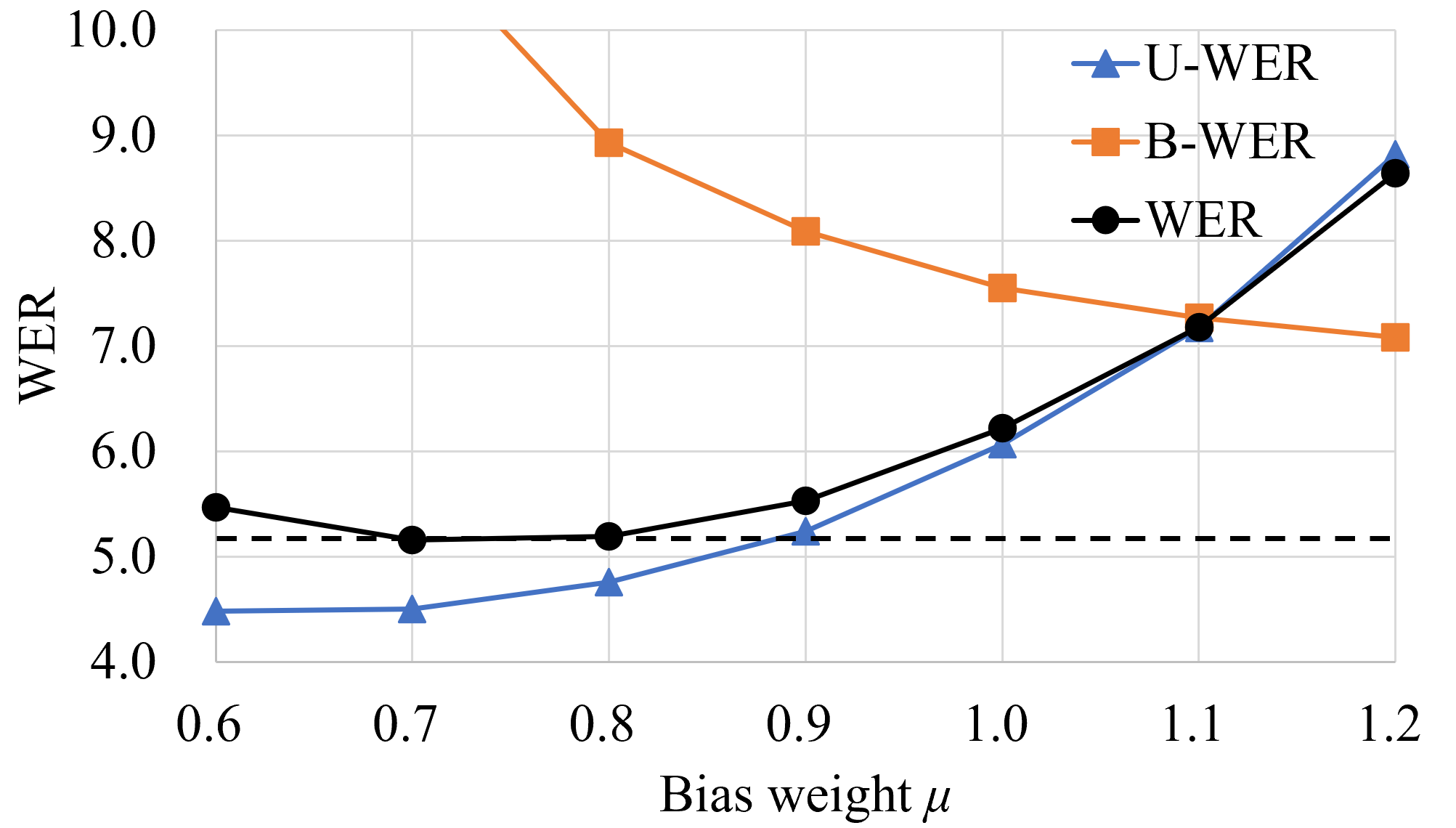} 
        \end{minipage}
    \vspace*{-3mm}
    \caption{Effect of the bias weight $\mu$.}
    \label{fig:weight}
\vspace*{-2mm}
\end{figure}

\subsection{Effect of bias weight during inference}
\label{sec:weight}

Figure \ref{fig:weight} shows the effect of the bias weights $\mu$ (Section \ref{sec:decoding}) on the WER, U-WER, and B-WER results for $N$ = 2000.
Increasing the bias weights $\mu$ improves the B-WER but deteriorates U-WER due to the tendency for overbiasing.
Under this experimental condition, there is a slight tendency for overbiasing when no bias weights are introduced; thus, when $\mu = 0.8$, the overbiasing can be suppressed.
The degree to which the model is biased depends on the target user domain; thus, we believe that this mechanism adjusting the bias weights easily during inference is effective.

\subsection{Validation on Japanese dataset}
\label{sec:japanese}

We validate the proposed method using our in-house Japanese dataset, comprising the Corpus of Spontaneous Japanese (581 h)~\cite{csj}, 181 h of Japanese speech from the database developed by the Advanced Telecommunications Research Institute International~\cite{KUREMATSU1990357}, and 93 h of our in-house Japanese speech data.
The CTC/attention-based system described in Section~\ref{sec:experimental condition} is used in this experiment.
Table \ref{castable} shows the results in terms of character error rate (CER), B-CER, and U-CER, with the bias list provided by our end users containing $N$ = 203 technical terms.
The proposed method significantly improves the B-CER with a slight degradation in U-CER, thereby resulting in the best overall CER.

\begin{table}[t]
\caption{Experimental results obtained on Japanese dataset.}
\vspace*{-7mm}
\label{castable}
\begin{center}
\resizebox {0.8\linewidth} {!} {
\begin{tabular}{@{}l|ccc}
\hline
Model & CER & U-CER & B-CER \\
\hline
Baseline & 9.85 & \textbf{8.17} & 21.76 \\
BPB beam search \cite{sudo2024_bias} & 9.67 & 9.20 & 13.16 \\
Intermediate DB \cite{shakeel2024_bias} & 9.28 & 8.23 & 16.93 \\
\textbf{Proposed} & \textbf{9.03} & 8.93 & \textbf{9.73} \\
\hline
\end{tabular}
}
\end{center}
\vspace*{-4mm}
\end{table}

\begin{figure}[t]
    \centering
        \begin{minipage}{0.45\textwidth}
            \includegraphics[width=\textwidth]{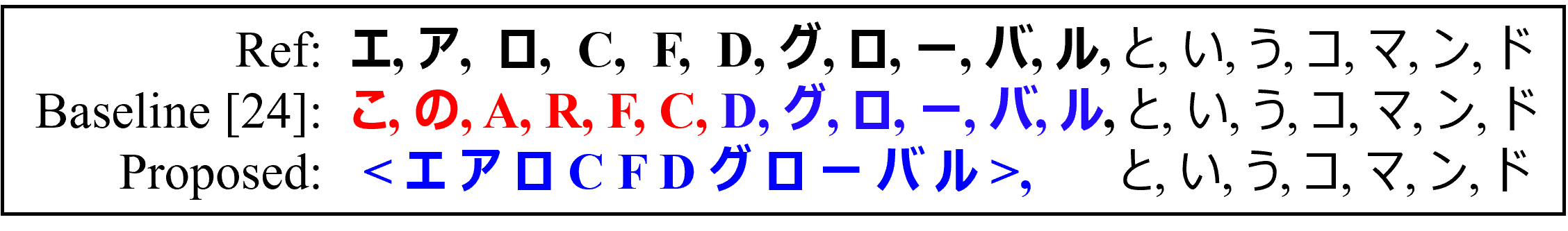} 
        \end{minipage}
        \centering
    \vspace*{-3mm}
    \caption{Typical inference example. The characters in boldface, red, and blue represent the bias phrases, incorrectly, and correctly recognized characters, respectively.
    }
    \label{fig:example}
\vspace*{-0mm}
\end{figure}

Figure \ref{fig:example} shows the typical inference results, where the characters in boldface, red, and blue represent the bias phrases, incorrectly, and correctly recognized characters, respectively.
As discussed in Section~\ref{sec:analysis}, the conventional DB method \cite{sudo2024_bias} struggles to capture the subword dependencies, especially in Japanese ASR, which operates at the character level, leading to longer subword sequences for bias phrases. 
In contrast, the proposed method avoids this problem by introducing the dynamic vocabulary where the bias token represents an entire bias phrase within a single token.

\subsection{Validation on the streaming RNN-T-based system}
\label{sec:transducer}

Table \ref{transducertable} shows the results of the streaming RNN-T-based systems with a bias list of size $N$ = 100, and 1000. 
The asterisk (*) indicates the use of external text data for model training (B1-B3).
We apply LM shallow fusion to the proposed method for fair comparison.
Note that bias tokens are decomposed into static subword token sequences before shallow fusion because the LM is not based on the dynamic vocabulary.
%Here, the hypotheses that are seven tokens shorter than the maximum length of the other hypotheses are removed, because the proposed method with the external LM tends not to generate long hypotheses, especially when dynamic tokens are included.
B1 and B2 incorporate the DB-based neural LM and unified speech-to-text representation (USTR), respectively \cite{Le2021ContextualizedSE,qiu2023improving}.

Consistent with the results from the offline CTC/attention-based system, the proposed method significantly improves the B-WER without relying on additional information, such as phonemes, with better overall WER than conventional DB methods (A1-2 vs. A3).
The conventional DB methods \cite{Le2021ContextualizedSE,qiu2023improving} considerably improve the B-WER by learning subword dependencies within the bias phrases using external text data (A1-2 vs. B1-2). In contrast, the proposed method eliminates this need by introducing the bias tokens. 
In addition, the proposed method with the external LM performed comparably to conventional DB methods (B3 vs. B1-2), although its main advantage is simplicity and high DB performance (B-WER) without relying on external text data.

\begin{table}[t]
\caption{WER results of the \textbf{streaming RNN-T-based} systems on Librispeech-960 test-clean (WER/B-WER). 
*850 M words of external text data is used for model training.
}
\vspace*{-7mm}
\label{transducertable}
\begin{center}
\resizebox {0.93\linewidth} {!} {
\begin{tabular}{@{}c|l|c|c}
\hline
ID & Model &  $N$ = 100 & $N$ = 1000 \\
\hline
A0 & Baseline (RNN-T) & 3.80 / 14.3 & 3.80 / 14.3 \\ %3.65
\hline
A1 & Trie-based DB \cite{Le2021ContextualizedSE} & 3.11 / 9.8 & 3.30 / 11.0 \\
A2 & Phoneme-based DB \cite{qiu2023improving} & 2.56 / 6.8 & 2.81 / 8.7 \\ 
A3 & \textbf{Proposed} & \textbf{2.43} / \textbf{3.1} & \textbf{2.66} / \textbf{3.5} \\  
\hline
B1 & A1+DB-LM*+FST \cite{Le2021ContextualizedSE} & 1.98 / 5.7 & \textbf{2.14} / 6.7 \\ 
B2 & A2+USTR*+FST \cite{qiu2023improving} & 2.06 / \textbf{2.0} & 2.16 / \textbf{2.5} \\ 
%B3 & \textbf{A3+LM}* & 2.12 / 2.9 & 2.92 / 3.4 \\ 
B3 & \textbf{A3+LM}* & \textbf{1.96} / 2.2 & 2.31 / 2.7 \\ % w/ length removal
\hline
\end{tabular}
}
\end{center}
\vspace*{-5.5mm}
\end{table}

\section{Conclusion}

In this paper, we present a simple but effective DB method that introduces a dynamic vocabulary where the bias tokens represent the bias phrases with the single entities. 
In addition, we introduce a bias weight to adjust the bias intensity during inference. 
The experimental results obtained by applying the proposed method to an offline CTC/attention-based system and a streaming RNN-T-based system demonstrate that it significantly improves bias phrase recognition on English and Japanese datasets.

%In future research, we plan to extend the proposed method to LMs to further improve the biasing performance.
%and multilingual models.

\clearpage

\bibliographystyle{IEEEbib}
%\bibliography{strings,refs}
\bibliography{mybib}

\end{document}